\documentclass[a4paper,12pt,draftclsnofoot,onecolumn]{IEEEtran}
\usepackage{amsmath}
\usepackage{graphicx}
\usepackage{caption}

\newtheorem{thm}{Theorem}[section]
\newtheorem{cor}[thm]{Corollary}
\newtheorem{lem}[thm]{Lemma}

\usepackage{amssymb}

\begin{document}

\title{From Graph Isoperimetric Inequality to Network Connectivity -- A New Approach}
\author{Qiang~Zhu$^{1}$,  %
\IEEEmembership{Member,~IEEE};
Fang~Ma$^{1}$;  %
\IEEEcompsocitemizethanks{$^{1}$ Q. Zhu, F. Ma and Weisheng.C are with the
School of Mathematics and Statistics, Xidian University, Xi'An, Shaanxi, China,
710071 \protect \
Email:~qiangzhu@ustc.edu, honestmafang@163.com,wshchen@126.com}
Guodong Guo$^2$, \IEEEmembership{Senior Member,~IEEE};
\IEEEcompsocitemizethanks{$^2$ G. Guo is with the Department of Computer Science
and Electrical Engineering, West Virginia University, Morgantown, WV 26506. \protect \
Email:~guodong.guo@mail.wvu.edu}
Dajin Wang$^3$;
\IEEEcompsocitemizethanks{$^3$ D. Wang is with the Department of
Computer~Science,~Montclair \protect State University, Montclair, NJ 07043. \protect \\ Email:~wangd@mail.montclair.edu}
Weisheng~Chen$^{1}$,  %
\IEEEmembership{Member,~IEEE}
\thanks{\newline
This work is supported by the National Natural Science Foundation of China
under Grant No. 11101322
}}
\maketitle

\begin{abstract}
We present a new, novel approach to obtaining a network's connectivity. More specifically,
we show that there exists a relationship between a network's {\it graph isoperimetric properties}
and its {\it conditional connectivity}.
A network's connectivity is the minimum number of nodes, whose
removal will cause the network disconnected. It is a basic and
important measure for the network's reliability, hence its
overall robustness. Several conditional connectivities have been proposed in
the past for the purpose of accurately reflecting various realistic
network situations, with {\it extra connectivity} being one such conditional
connectivity. In this paper, we will use isoperimetric properties
of the hypercube network to obtain its extra connectivity.
The result of the paper for the first time establishes a relationship between
the age-old isoperimetric problem and network connectivity.
\end{abstract}


\section{Introduction}
With the increase of the number of processors in multiprocessor computer systems, the possibility of some nodes failing/malfunctioning increases as well.
The overall system reliability is therefore a key issue in the design, implementation, and maintenance of large multiprocessor systems.
There are two basic criteria in evaluating the reliability of multiprocessor systems. One is to determine if a certain structure can be
embedded into the remaining healthy system. The other is to determine whether a fault-free communication path exists between any two fault-free nodes.
We focus on the latter in this paper.

A multiprocessor system at the system-level can be modeled with an undirected graph $G(V, E)$. Each vertex (or node) in $V(G)$ represents a processor in the multiprocessor system, and each edge in $E(G)$ represents a communication link between two processors.  A vertex cut $S$ (resp. an edge cut $S$) of a graph $G$ is a vertex subset $S\subset V(G)$ (resp. an edge subset $S\subset E(G)$) such that $G-S$ is disconnected. The connectivity (resp. edge connectivity) of a graph $G$ is the cardinality of a minimum vertex cut (resp. edge cut) of $G$.
Connectivity (resp. edge connectivity) has been used as a traditional measure to evaluate the fault tolerance ability of multiprocessor systems.
However, it has shown some deficiencies as a measure for fault tolerance. On one hand, as surveyed in \cite{Boesch1986-p240-246},
it cannot correctly reflect different situations of disconnected graphs when removing a vertex cut, which will render inaccuracy for some applications;
on the other hand, for many interconnection networks, the probability that all vertices in a minimum vertex cut fail at the same time is quite small.
So the classical definition of connectivity may have over-pessimistically underestimated many networks' reliability \cite{Esfahanian1989-p1586-1591}.

Motivated by the above-mentioned shortcomings, Harary \cite{Harary1983p347-357} introduced the concept of conditional connectivity
by requiring that the disconnected components of $G-F$ have certain properties.
Restricted connectivity, super connectivity and $h$-extra connectivity are examples of conditional connectivity,
proposed by  A.H.Esfahanian and S.L.Hakimi\cite{Esfahanian1988-p195-199},
D. Bauer et. al.\cite{Bauer1981p89-98,Boesch1986-p240-246},
J. F$\grave{a}$brega and M.A. Fiol\cite{F`abrega1994-p163-170}, respectively.
All these connectivities require some properties of the disconnected components, or have some restrictions on the faulty sets. Thus they are more refined measures of reliability for multiprocessor systems.
The restricted connectivity,  super connectivity and extra connectivity of many interconnection networks have been explored \cite{Balbuena2007-p2444-2455,  Balbuena2008-p1985-1993,  Balbuena2009-p1586-1591,  Balbuena2008-p2827-2834,  Chen2009-p1202-1205,  Chen2007p1848-1855, Liu2009-p655-659,  Volkmann2007-p234-239,  Yang2009-p887-891, Zhu2006-p111-121,  Day1994p31-38,  Wang2008-p587-596,  Wang2004-p199-205,  WangM2002-p205-214,  Ma2008-p59-63,  Yang2009-p887-891,  Lin2005p257-271,  Xu2005p191-195,  Xu2010p25-32,  Li1999p157-159,  Meng2002p183-193,  Soneoka1992p511-523,  Guo2010p659-661,  Lu2007p152-157,  Ma2011p360-364,  Zhu2007p1782-1788,  Zhu2006-p111-121,  Li1998p61-65,  Zhu2008-p1069-1077,  Hong2012p1-11, Yang2009-p887-891,  zhou2009study,  hsieh2012extraconnectivity, chang20133,  zhu2013reliability, Yang2012p1335-1339,  Zhou2012p2152-2164}.

When faults occur in an interconnection network, it may become disconnected. But if the disconnected network has a very large component and the remaining small components have very few vertices in total, its performance will not degrade dramatically. This is much preferred than having a disconnected graph without any large components. This phenomenon has been studied in \cite{Chen2007p1848-1855, Cheng2000p97-112,  Cheng2002p165-169, Cheng2007p4877-4882}.

The hypercube \cite{Saad1988p867-872} is a well-known interconnection network for multiprocessor computers. It possesses many attractive properties. The restricted connectivity, super connectivity, and extra connectivity of hypercubes have been studied  in \cite{chang20133, Hong2012p1-11, Yang2009-p887-891, Yang2012p1335-1339, Zhu2006-p111-121}.

The age-old isoperimetric problem dates back to ancient literature, and in its original form is about finding the largest possible area with
a given boundary length. In graph theory, an isoperimetric inequality is a lower bound for the size of the boundary in terms of the order of subgraph.
The isoperimetric problem of many graphs have been studied \cite{alon1985lambda, Harper1966p385-393, kolesnik2014lower, subramanya2009bounds}.
However to the best of our knowledge, the isoperimetric problem for graphs has never been related to network connectivity/reliabilty in the past.

In this paper, we will for the first time establish a relationship between
network connectivity and graph isoperimetric problem.
Opening a new direction in the study of interconnection networks,
we will use the isoperimetric properties of the hypercube to study its reliability. More specifically,
we will show that
if the number of removed vertices is less than the $h$-minimum vertex boundary number of hypercube $Q_n$,
$1\le h \le 3n-6$, then there must exist a large component, and the total number of vertices in the remaining
small components is upper-bounded by a function of $h$.
We will then prove that when
\begin{itemize}
\item $1\le h \le n-3 $ and $n\ge 5$; or
\item $n+2\le h \le 2n-4$ and $n\ge 7$; or
\item $2n+1 \le h \le 3n-6$ and $n\ge9$.
\end{itemize}
the hypercube $Q_n$'s $(h-1)$-extra connectivity is equal to its minimum $h$-vertex boundary number;
when $n-2\le h \le n+1$ and $n\ge 5$, $Q_n$'s $(h-1)$-extra connectivity is equal to its minimum $(n-2)$-vertex boundary number;
when $2n-3\le h \le 2n$ and $n\ge 7$, $Q_n$'s $(h-1)$-extra connectivity is equal to its minimum $(2n-3)$-vertex boundary number. \\

The rest of this paper is organized as follows. Section II provides preliminaries, and introduces terminology and useful lemmas.
In section III, we use the results on the isoperimetric problems for the hypercube to explore the structure of faulty hypercube, and
determine its extra connectivity. Section IV summarizes the paper with concluding remarks.



\section{Preliminaries and terminologies}
For all terminologies and notations not defined in this section,  we follow \cite{J.A.Bondy1976}. Let $G=(V, E)$ be a simple undirected graph. For a  vertex subset or a subgraph $H$ of $G$,  the vertex boundary $N_G(H)$ of $H$ is the set of vertices not in $V(H)$ joined to some vertices in $V(H)$. We use $C_{G}(H)$ to denote the set $N_{G}(H)\bigcup V(H)$.  The vertex boundary number of $H$ is the number of vertices in $N_G(H)$, denoted by $b_v(H;G)$.  The minimum $k$-boundary number of $G$ is defined as the minimum boundary number of all its subgraphs  with order $k$,  denoted by $b_{v}(k;G )$. 
Given a vertex subset $S$,   we use  $[S]$ to denote the induced subgraph of $S$ in $G$ and  $G-S$ to denote the induced subgraph $[V(G) -S]$. The $cn$-number of $G$ is defined as the maximum number of common neighbors shared by a pair of different vertices in $G$, denoted by $cn(G)$\cite{xuzhu2005}.

Given a graph $G$, a vertex subset $S$ is called a vertex cut if $G-S$ is disconnected or trivial. The connectivity of $G$ is the cardinality of the smallest vertex cut, denoted by $\kappa(G)$. Given a graph $G$, a vertex cut $F$ is called a super vertex cut if  each connected component of $G-F$ has at least 2 vertices. The super connectivity  $\kappa_1(G)$ of $G$ is defined as the cardinality of the smallest super vertex cut of $G$, if exists. Given a graph $G$ and a non-negative integer $h$,  a vertex cut $T$ of $G$ is called an $h$-extra vertex cut if  each component of $G-T$ has at least $h+1$ vertices. The $h$-extra connectivity  $\kappa_h(G)$ of $G$ is defined as the cardinality of the smallest $h$-extra vertex cut of $G$, if exists. By the above definitions, the 1-extra connectivity is the super connectivity. So extra connectivity is a generalization of the widely stuided super connectivity and may provide more accurate measures of the reliability of some interconnection networks.

The hypercubes are the most famous and widely studied class of interconnection networks. The vertices in an $n$-dimensional hypercube can be labelled by $n$-bit binary strings. Two vertices are adjacent if and only if they differ in exactly one bit position.
Let $\oplus$ denotes the binary operation exclusive or.
For a vertex $u=u_1 u_2 \cdots u_n$,   let  $u^{i}=u_1 \cdots u_{i-1} \overline{u_i} u_{i+1}\cdots u_n$ where $\overline{u_i}= {u_i} \oplus 1$. We call $u^{i}$  the $i$-th neighbor of $u$. Similarly, $u^{i j}$  the $j$-th neighbor of $u^{i}$ and $u^{i j k}$  the $k$-th neighbor of $u^{i j}$. For an edge  $e=(u, u^{i})$, we call $e$ an $i$-th edge.
Thus an $n$-dimensional hypercube has $2^n$ vertices,  each vertex has $n$ neighboring vertices.

 Given an $i$,  $1\le i \le n$,  let $S_i^{0}=\{u_1u_2\cdots u_n| u_i= 0\}$,  $S_i^{1}=\{u_1u_2\cdots u_n| u_i= 1\}$. By the definition of hypercubes,  the induced subgraph of $S_i^{0}$ and $S_i^{1}$ are both isomorphic to an $(n-1)$-dimensional hypercube. Furthermore,  there exists a perfect matching $M_i$ between $S_i^{0}$ and $S_i^{1}$ in $Q_n$. All $i$-th edges belong to $M_i$ and $M_i$ contains only $i$-th edges. We call this  a decomposition of an $n$-dimensional hypercube along the $i$-th dimension,  denoted by $Q_n =G(Q_{n-1}^{0},  Q_{n-1}^{1}; M_i )$. Given the decomposition, we call the edges in $M_i$ inter edges while other edges inner edges. Given the decomposition of an $n$-cube along the $i$-th dimension, we call the $i$-th neighbor of any vertex $u$ the pair vertex of $u$ in the decomposition.

%
%
%


By the definition of the hypercube, it's easy to know that the  connectivity of an $n$-dimensional hypercube $Q_n$ is $n$ \cite{Harary1988survey}.

\begin{lem}\label{connectivity}
Given an $n$-dimensional hypercube $Q_n$, $\kappa(Q_n)=n$.
\end{lem}

The $cn$-number of hypercubes are first explored in \cite{xuzhu2005}.
\begin{lem}\cite{xuzhu2005} \label{cnnumber}
Given an $n$-dimensional hypercube $Q_n$ with $n\ge 2$, any pair of different  vertices in $V(Q_n)$  have exactly two common neighbors if they have any.
\end{lem}

In 1966, Harper et. al. determined the minimum $m$-vertex boundary number of $Q_n$.

\begin{lem}\label{qboudary}\cite{Harper1966p385-393}
Every integer $m$,  $1\le m \le 2^n -1$,  has a unique representation in the form
$$ m= \sum_{i=r+1}^{n}\binom{n}{i} + m^{'},  \quad 0<  m^{'} \le \binom{n}{r},  $$
$$m^{'}=\sum_{j=s}^{r}\binom{m_j}{j},  \quad 1\le s\le m_s< m_{s+1} < \ldots < m_{r}. $$

Then
   $$b_v(m; Q_n)=\binom{n}{r}-m^{'}+\sum_{j=s}^{r}\binom{m_j}{j-1}.  $$
\end{lem}

\section{Minimum boundary number of hypercubes}

By lemma \ref{qboudary},  each integer $m\le 2^n -1$ has a unique representation  and the minimum $m$-vertex boundary number of $Q_n$ can be obtained by this representation. For an integer $m$ between 1 and $2^{n-1}$,  we use $r(m)$ (resp. $s(m)$) to denote the $r$ (resp. $s$ ) in the above unique representation of $m$.
%
%

By Lemma \ref{qboudary}, the following lemma can be obtained.
\begin{lem}\label{repr}
Let $m$, $M$ be two different integers between $1$ and $2^n -1$, suppose the unique expression of $m$, $M$ are as follows:
$$ m= \sum_{i=r_m+1}^{n}\binom{n}{i} + m^{'},  \quad 0<  m^{'} \le \binom{n}{r_m}; M= \sum_{i=r_M+1}^{n}\binom{n}{i} + M^{'},  \quad 0<  M^{'} \le \binom{n}{r_M} $$

$$m^{'}=\sum_{j=s_m}^{r_m}\binom{m_j}{j},  \quad 1\le s_m\le m_s< m_{s+1} < \ldots < m_{r_m} $$

$$M^{'}=\sum_{j=s_M}^{r_M}\binom{M_j}{j},  \quad 1\le s\le M_s< M_{s+1} < \ldots < M_{r_M} $$

Then $m<M$ if and only if one of the following conditions holds:

(1) $r_m > r_M$.

(2) $r_m =r_M =r$, $m_j < M_j$, $m_k =M_k $ $k= j+1, \cdots, r$.

\end{lem}

By Lemma \ref{repr}, it's easy to determine the explicit  minimum $m$-boundary number of $Q_n$. As shown in the flowing corollary,
 which gives the explicit expression of the minimum $m$-boundary number of $Q_n$ when $1\le m \le 6n-15$.

\begin{cor} \label{boundary}
\begin{displaymath}
b_{v}(m;Q_{n}) =\left\{\begin{array}{ll}
n ,    & \textrm{ $m=1$}\\
- \frac{m^2}{2} +(n-\frac{1}{2})m+ 1 ,     & \textrm{$ 2\leqslant m \leqslant n+1 $} \\
-\frac{m^2}{2}+(2n-\frac{3}{2})m-n^2+2 ,   & \textrm{$ n+2\leq m\leq 2n-1$ } \\
-\frac{m^2}{2}+(3n-\frac{7}{2})m-3n^2+4n+2,    & \textrm{$2n \leq m \leq  3n-3$ }\\
-\frac{m^2}{2}+(4n-\frac{13}{2})m-6n^2+15n-4,    & \textrm{$ 3n-2 \leq m \leq 4n-6 $}\\
-\frac{m^2}{2}+(5n-\frac{21}{2})m-10n^2+36n-24,    & \textrm{$ 4n-5 \leq m \leq 5n-10 $}\\
-\frac{m^2}{2}+(6n-\frac{31}{2})m-15n^2+70n-69.    & \textrm{$ 5n-9 \leq m \leq 6n-15 $}

\end{array} \right.
\end{displaymath}
\end{cor}

$\mathbf{Proof:}$  It's clear that $r(m)\leq n$ for any integer $m$.

1) When  $r(m)=n$. By lemma \ref{qboudary},   $ m=  m^{'} =1 $, $s(m)=n$, $m_{n}=n$.  Thus $b_v(1; Q_n)=n$.

2) When $r(m)=n-1$, $m = 1 + m^{'}$, $0<m^{'}\leq n$.
 So $2\le m\le n+1$.

  When $2\leq m \leq n$, since in the unique representation of $m^{'}=\sum_{j=s}^{n-1}\binom{m_j}{j},  \quad 1\le s\le m_s< m_{s+1} < \ldots < m_{n-1}$,  $m_{n-1}\geq n-1$.  If   $m_{n-1}=n$, $ m^{'}\geq \binom{m_{n-1}}{n-1} =n$ which contradicts to $m^{'}\le n-1$. So $m_{n-1}=n-1$.  Thus $m_j=j$ for $s\le j\le n-1$, $m^{'}=n-1-(s-1)$, $s=n-m^{'}= n-m+1$. So $b_v(m; Q_n)= \binom{n}{n-1}-(m-1)+((n-1)+(n-2)+\cdots +(n-m+1)= - \frac{m^2}{2} +(n-\frac{1}{2})m+ 1$.

 When $m=n+1$,    $m^{'}= m- \binom{n}{n}=n$.  Since in the unique representation of $m^{'}=\sum_{j=s}^{n-1}\binom{m_j}{j},  \quad 1\le s\le m_s< m_{s+1} < \ldots < m_{n-1}$,  $m_{n-1}=n$ and  $s= n-1$.  So $b_v(n+1; Q_n)= \binom{n}{n-1}-n +\binom{n}{n-2}= \binom{n}{n-2}= - \frac{(n+1)^2}{2} +(n-\frac{1}{2})(n+1)+ 1$.

  So $b_v(m; Q_n)=  - \frac{m^2}{2} +(n-\frac{1}{2})m+ 1$ for $ 2\leqslant m \leqslant n+1 $.

3)When $r(m)= n-2$, $m = 1+ n + m^{'}$, $0< m^{'}\leq  \binom{n}{n-2}$.
Thus $n+2\leq m\leq \frac{n(n+1)}{2}+1$.   Since in the unique representation of $m^{'}=\sum_{j=s}^{n-2}\binom{m_j}{j},  \quad 1\le s\le m_s< m_{s+1} < \ldots < m_{n-2}$,  $m_{n-2} \geq n-2$,  then we will discuss according to the following cases:

Case~1: $m_{n-2}= n-2$, then   $m_j=j$ for $1\leq s\le j\le n-2$. Since $m^{'}=\sum_{j=s}^{n-2}\binom{m_j}{j}=n-2-s+1$ for $ 1\leq s\le j\le n-2$,  then $1\leq m^{'} \leq n-2$,    $n+2\leq m\leq 2n-1$.  $s=n-2+1-m^{'}=2n-m$.  So $b_v(m; Q_n)= \binom{n}{n-2}-(m-n-1)+((n-2)+(n-3)+\cdots +(2n-m)=-\frac{m^2}{2}+(2n-\frac{3}{2})m-n^2+2$.

Case~2: $m_{n-2}= n-1$ and $m_{n-3}= n-3$,  then $m_j=j$ for $s\le j\le n-3$. Since $m^{'}=\sum_{j=s}^{n-2}\binom{m_j}{j}=\binom {n-1}{n-2}+(n-3-s+1)$ for $1\leq s\le j\le n-2$,  then $n-1\leq m^{'} \leq 2n-4$,    $m=n+1+m^{'}$,   $2n \leq m\leq 3n-3$.  $s=2n-3-m^{'}=3n-m-2 $,  So $ b_v(m; Q_n)= \binom{n}{n-2}-(m-n-1)+\binom{n-1}{n-3}+((n-3)+(n-4)+\cdots +(3n-m-2))=-\frac{m^2}{2}+(3n-\frac{7}{2})m-3n^2+4n+2$.

Case~3: $m_{n-2}= n-1$,  $m_{n-3}= n-2$ and  $m_{n-4}= n-4$,  then $m_{j}= j$ for $s\leq j\leq n-4$. Since $m^{'}=\sum_{j=s}^{n-2}\binom{m_j}{j}=\binom {n-1}{n-2}+\binom {n-2}{n-3}+(n-4-s+1)=3n-s-6$  for $1\leq s\le j\le n-3$,  $2n-3 \leq m^{'} \leq 3n-7$,  $3n-2 \leq m\leq 4n-6$,  $s=4n-m-5$. So $b_v(m; Q_n)= \binom{n}{n-2}-(m-n-1)+\binom{n-1}{n-3} +\binom{n-2}{n-4}+((n-4)+(n-5)+\cdots +(4n-m-5)=-\frac{m^2}{2}+(4n-\frac{13}{2})m-6n^2+15n-4$.

Case~4: $m_{n-2}= n-1$,  $m_{n-3}= n-2$, $m_{n-4}= n-3$ and $m_{n-5}= n-5$,  then $m_{j}= j$ for $s\leq j\leq n-5$. Since $m^{'}=\sum_{j=s}^{n-2}\binom{m_j}{j}=\binom {n-1}{n-2}+\binom {n-2}{n-3}+\binom {n-3}{n-4}+(n-5-s+1)=4n-s-10$  for $1\leq s\le j\le n-4$,  $3n-6 \leq m^{'} \leq 4n-11$,  $4n-5 \leq m\leq 5n-10$,  $s=5n-m-9$. So $b_v(m; Q_n)= \binom{n}{n-2}-(m-n-1)+\binom{n-1}{n-3} +\binom{n-2}{n-4}+\binom{n-3}{n-5}+((n-5)+(n-6)+\cdots +(5n-m-9)=-\frac{m^2}{2}+(5n-\frac{21}{2})m-10n^2+36n-24$.

Case ~5: $m_{n-2}= n-1$,  $m_{n-3}= n-2$, $m_{n-4}= n-3$, $m_{n-5}= n-4$  and $m_{n-6}= n-6$,  then $m_{j}= j$ for $s\leq j\leq n-6$. Since $m^{'}=\sum_{j=s}^{n-2}\binom{m_j}{j}=\binom {n-1}{n-2}+\binom {n-2}{n-3}+\binom {n-3}{n-4}+ \binom {n-4}{n-5}+(n-6-s+1)=5n-s-15$  for $1\leq s\le j\le n-5$,  $4n-10 \leq m^{'} \leq 5n-16$,  $5n-9 \leq m\leq 6n-15$,  $s=6n-m-14$. So $b_v(m; Q_n)= \binom{n}{n-2}-(m-n-1)+\binom{n-1}{n-3} +\binom{n-2}{n-4}+\binom{n-3}{n-5}+\binom{n-4}{n-6}+((n-6)+(n-7)+\cdots +(6n-m-14)=-\frac{m^2}{2}+(6n-\frac{31}{2})m-15n^2+70n-69$.

\hfill\rule{1mm}{2mm}

By Corollary \ref{boundary}, the following  results are immediate.

\begin{lem}\label{boundary2}
 Given an $n$-dimensional hypercube $Q_n$, then

$b_{v}(i*n-1-\frac{i*(i-1)}{2};Q_{n})= b_{v}(i*n-\frac{i*(i-1)}{2};Q_{n})= b_{v}(i*n-2-\frac{i*(i-1)}{2};Q_{n})+1=b_{v}(i*n+1-\frac{i*(i-1)}{2};Q_{n})+1$ when $1\le i \le 5$,

$b_{v}(m;Q_{n})< b_{v}(m+1;Q_{n})$ for $ 1\le m \le n-2$, $n+1\le m \le 2n-3$, $2n\le m \le 3n-5$, $3n-2\le m \le 4n-8$, $4n-5 \le m \le 5n-12$, $5n-9 \le m \le 6n-17$.

$b_{v}(i*n+2-\frac{i*(i-1)}{2};Q_{n})> b_{v}(i*n-\frac{i*(i-1)}{2};Q_{n})$ when $1\le i \le 5$ and $n-i \ge 4$.
\end{lem}

%
%

According to the above Lemma,  the following results  can be obtained.
\begin{lem}\label{lowhigh} Given an $n$-dimensional hypercube $Q_n$ with $n\ge 5$, we have:

$b_{v}(h ;Q_{n})-b_{v}(h-1; Q_{n-1})=n-1$ for $ 1 \leq h \leq n+1$.

$b_{v}(h ;Q_{n})-b_{v}(h-1 ;Q_{n-1})=h-2 $ for $ n+2 \leq h \leq 2n-1$.

\end{lem}

%
The following lemma shows that there exists a large component in $Q_n -S$ when $|S|< b_{v}(h;Q_{n})$ for $h\le n-2$.

\begin{lem} \label{fs}
Given an $n$-dimensioanl hypercube $Q_n$  where $n\geq 5$, for any vertex subset  $S$ of $Q_{n}$ with  $|S|< b_{v}(h;Q_{n})$ for $1 \le h\leq n-2$,   $Q_{n}-S$ has a large component and all the remaining small components have at most $h-1$ vertices in total.

\end{lem}
$\mathbf{Proof:}$  We use induction to prove this.

$\mathbf{(i)}$. When $h=1$,  $b_{v}(1;Q_{n})=n=\kappa(Q_n)$ by Lemma \ref{cnnumber}.  Thus $Q_{n}-S$ is connected when $|S|<b_{v}(1;Q_{n})$,   the proposition holds.

$\mathbf{(ii)}$. Suppose the proposition holds for $h-1$ where $h\ge 2$,  in the following we use contradiction to prove that it also holds for $ h$. Suppose not,  let $C_{1}, C_{2}, \cdots , C_{m}$ be all the components of $Q_{n}-S$ and $|V(C_{1})| \leq |V(C_{2})|\leq\cdots \leq |V(C_{m})|$,  then $m\geq 2$ and  $\sum _{i=1}^{m-1}(|V(C_{i})|)\geq h$. Let $ B=\bigcup _{i=1}^{m-1}V(C_{i}) $,  $\forall\  u=u_{1}u_{2} \cdots u_{n}$,  $ v=v_{1}v_{2} \cdots v_{n}$,  $u, v\in B$,  $ u \neq v $,  then $\exists$ $ i$,   s.t.$u_{i}\neq v_{i}$,  let $Q_{n}=G(Q^{0}_{n-1}, Q^{1}_{n-1};M_i)$ be a decomposition of $Q_{n}$ along the $i$-th dimension,  then $u, v$ don't belong to the same subcube.

Let $S_{i}=S\bigcap V(Q_{n-1}^{i})$,  $i=0, 1$,   then $|S_{0}|+|S_{1}|=|S|< b_{v}(h; Q_n)$. By Lemma \ref{lowhigh},    $b_{v}(h ;Q_{n})-b_{v}(h-1;Q_{n-1})=n-1$.  It can be verified that $n-1\leq b_{v}(h-1;Q_{n-1}) $ for $ n\geq 2$.  So at most one of $|S_{0}|$ and $|S_{1}|$ can be not less than $b_{v}(h-1;Q_{n-1})$.

$\mathbf{Case ~1}$. $|S_{0}|< b_{v}(h-1;Q_{n-1})$ and $|S_{1}|< b_{v}(h-1;Q_{n-1})$.

  By the induction
hypothesis,   $Q^{0}_{n-1}- S_{0}$ $(resp.Q^{1}_{n-1}- S_{1})$ has a large component $C_{0} (resp.C_{1})$,  and all the small components have at most $h-2$ vertices in total. Since there exists a perfect matching $M$ between $Q^{0}_{n-1}$ and $ Q^{1}_{n-1}$ and has at least $2^{n-1}-|S|-2(h-2)> 0$ edges between $C_{0}$ and $C_{1}$ in $Q_{n}-S$, thus $C_{0}$ and $ C_{1}$ are connect to each other in $Q_{n}-S$. So  there exist a large component $C$ in $Q_{n}-S$,  and all the small components have at most $2(h-2)\leq 2(n-4)=2n-8$ vertices in total. Let $A$ denotes the union of all the vertex sets of the small components in $Q_{n}-S$,  if $|A|\leq h-1$,  then we are done. Suppose not,  then $h\leq |A|\leq 2n-8$.  Since $N_{Q_{n}}(A)\subset S$,  $|S|\geq |N_{Q_{n}}(A)|\geq b_{v}(|A|;Q_{n})$. By Lemma \ref{boundary2},  $b_{v}(h;Q_{n})<b_{v}(l;Q_{n})$ for $n\ge 5$, $h\leq n-2$ and $ h<l\le 2n-8$. So $|S|\geq b_{v}(|A|;Q_{n})\geq b_{v}(h;Q_{n})$,  this is a contradiction to $|S|< b_{v}(h;Q_{n})$.

 $\mathbf{Case ~2}$. $|S_{0}|\geq b_{v}(h-1, Q_{h-1})$ or $|S_{1}|\geq b_{v}(h-1, Q_{n-1})$.

  Without loss of generality,  suppose $|S_{1}|\geq b_{v}(h-1;Q_{n-1})$,  then $|S_{0}|=|S|-|S_{1}|\leq n-2$,   $Q_{n}^{0}-S_{0}$ is connected. There exists a large component in $Q_{n}-S$ which  contains all vertices in $Q_{n}^{0}-S_{0}$,  so there is no vertex of the small components in $Q_{n}^{0}-S_{0}$,  which contradicts to the decomposition of $Q_{n}$. \hfill\rule{1mm}{2mm}

\begin{lem}\label{fs2}
Let $S$ be a vertex set in $Q_{n}$  with $n\geq 5$ and $b_{v}(n-2;Q_{n})\le |S|<b_{v}(n-1;Q_{n})$,  then $Q_{n}-S$ has a large component and all the remaining small components have at most $n+1$ vertices in total.
\end{lem}
$\mathbf{Proof:}$  Let $S_{i}=S\bigcap V(Q^{i}_{n-1})$,  $i=0,  1$. According  to Corollary 2.4,   $b_{v}(n-2;Q_{n})=\frac{n^2-n}{2}$,  $b_{v}(n-1;Q_{n})=\frac{n^2-n}{2}+1$,  so $|S|=\frac{n^2-n}{2}$. Since $b_{v}(n-3;Q_{n-1})+b_{v}(1;Q_{n-1})=b_{v}(n-2;Q_{n})$,   it can be verified that $b_{v}(1;Q_{n-1})\leq b_{v}(n-3;Q_{n-1}) $ for $ n\geq 4$, so at most one of $|S_{0}|$ and $|S_{1}|$ can be not less than $ b_{v}(n-3, Q_{n-1})$.

$\mathbf{Case ~1}$. $|S_{0}|< b_{v}(n-3;Q_{n-1})$ and $|S_{1}|< b_{v}(n-3;Q_{n-1})$.

By Lemma 3.5,  $Q^{0}_{n-1}- S_{0}$ $(resp.Q^{1}_{n-1}- S_{1})$ has a large component $C_{0} (resp.C_{1})$,  and all the small components have at most $n-4$ vertices in total. And similarly as Lemma $3.5$,   $C_{0}$ and $ C_{1}$ can be proved to be connected to each other in $Q_{n}-S$,  So there exists a large component $C$ in $Q_{n}-S$.   Let $A$ denotes the union of all the vertex sets of the small components in $Q_{n}-S$,  if $A\leq n+1$,  then we are done. Suppose not,  we have $n+2\leq |A|\leq 2n-8$. Similarly,  $|S|\geq |N_{Q_{n}}(A)|\geq b_{v}(|A|;Q_{n})\geq b_{v}(n+2;Q_{n})> \frac{n^2-n}{2} $ by Lemma \ref{boundary2},  which contradicts to   $|S|=\frac{n^2-n}{2}$.

 $\mathbf{Case ~2}$.  $|S_{0}|\geq b_{v}(n-3;Q_{n-1})$ or $|S_{1}|\geq b_{v}(n-3;Q_{n-1})$.

Without loss of generality,  we assume $|S_{1}|\geq b_{v}(n-3;Q_{n-1})$,  then $|S_0|<n-1= b_{v}(1;Q_{n-1})$, so  $Q_{n}^{0}-S_{0}$ is connected. So there is no vertex of the small components in $Q_{n}^{0}-S_{0}$,  which contradicts to our construction of the decomposition of $Q_{n}$. \hfill\rule{1mm}{2mm}

\begin{lem}\label{fs3}
Let $S$ be a vertex set in $Q_{n}$,  $ b_{v}(n-1;Q_{n})\leq |S|< b_{v}(h;Q_{n})$ for $n\geq 7$ and $ n+2\leq h\leq 2n-3$,  then $Q_{n}-S$ has a large component and all the small components have at most $h-1$ vertices in total.
\end{lem}
$\mathbf{Proof:}$  We use induction to prove it.

$\mathbf{(i)}$. When $h=n+2$. Since $b_{v}(n+2;Q_{n})-b_{v}(n-2;Q_{n-1})=2n-5$ and  $2n-5\leq b_{v}(n-2;Q_{n-1})$,   then at most one of $S_0$ and $S_1$ can be not less than $b_{v}(n-2;Q_{n-1})$.

$\mathbf{Case ~1.1}$. $|S_{0}| < b_{v}(n-2;Q_{n-1})$ and $|S_{1}| < b_{v}(n-2;Q_{n-1}) $.

By Lemma \ref{fs} and Lemma \ref{fs2},  $Q_{n-1}^{0}-S_{0}(resp. Q_{n-1}^{1}-S_{1})$ has a large component and all small components have at most $n$ vertices in total. Similarly as the proof of Lemma \ref{fs2},   the two large components of $Q_{n-1}^{0}-S_{0}$ and $Q_{n-1}^{1}-S_{1}$ are connected in $Q_n -S$. Let $A$ denotes the union of  the vertex sets of all the small components in $Q_n -S$,   then $|A|\leq n+n$. If $|A|\leq n+1$,  we are done. Suppose not, then $n+2 \leq |A|\leq 2n$.   According to Lemma \ref{boundary2},  we have $b_{v}(n+2;Q_{n})\leq b_{v}(l;Q_{n})$  for $n+2\leq l\leq 2n$.  So $|S|\geq |N_{Q_{n}}(A)|\geq b_{v}(|A|;Q_{n})\geq  b_{v}(n+2;Q_{n})$,  this is a contradiction to $|S|< b_{v}(n+2;Q_{n})$.

 $\mathbf{Case ~1.2}$. $|S_{0}|\geq b_{v}(n-2;Q_{n-1})$ or $|S_{1}| \geq b_{v}(n-2;Q_{n-1})$.

Without loss of generality,  we assume $|S_{0}|\geq b_{v}(n-2;Q_{n-1})$,  then $|S_{1}|\leq 2n-6< b_{v}(2;Q_{n-1})=2n-4$,  so  $Q_{n-1}^{1}-S_{1}$ has a large component and the small components have at most $1$ vertex in total. By the construction of the decomposition,  the small components have at least $1$ vertex in $Q_{n-1}^{1}-S_{1}$,  so the small components has exactly $1$ vertex in $Q_{n-1}^{1}-S_{1}$,  we use $u$ to denote it. Since there exists a perfect matching between $Q_{n-1}^{0}$ and $Q_{n-1}^{1}$ in $Q_n$, the vertices of $Q_{n-1}^{0}-S_0$ whose pair vertex is not in $S_1\cup\{u\}$ must be in the large component of $Q_n -S$.
Let $A$ denotes the union of all the vertex sets of the small components in $Q_{n}-S$,  then $|A|\le |S_1|+2\le 2n-4$.
if $|A|\leq n+1$,  we are done. Suppose not,we assume $|A|\geq n+2$. Then $2n-5\ge |A|\geq n+2$. Similar to the proof of Case 1. of Lemma 3.5, a contradiction can be obtained.


$\mathbf{(ii)}$. Suppose the proposition holds for $h-1$ where $h\geq n+3$,  in the following we will prove it also holds for $h$. Suppose not,  let $C_{1}, C_{2}, \cdots, C_{m}$ be all the components in $Q_{n}-S$;  that is,  $m\geq 2$  and $\sum_{i=1}^{n-1}|V(C_{i})|\geq h$.  Let $ B=\bigcup _{i=1}^{m-1}V(C_{i}) $, $u, v$ be two different vertices in $B$. Suppose they differ in the $i$-th dimension,
  let $Q_{n}=G(Q^{0}_{n-1}, Q^{1}_{n-1};M)$ be a decomposition of $Q_{n}$ along the $i$-th dimension.  Then $u$ and $ v$ don't belong to the same subcube.

Since $b_{v}(h-1;Q_{n-1})+ (h-2)=b_{v}(h;Q_{n})$,  it can be verified that $h-2\leq b_{v}(h-1;Q_{n-1}) $ when $ n\geq 7$ and $h\geq n+3$. So at most one of $|S_{0}|$ and $|S_{1}|$ can be not  less than $ b_{v}(h-1;Q_{n-1})$.

$\mathbf{Case ~2.1}$. $|S_{0}|<b_{v}(h-1;Q_{n-1})$ and $|S_{1}|< b_{v}(h-1;Q_{n-1})$.

By the induction hypothesis, $Q_{n-1}^{0}-S_{0}$(resp. $Q_{n-1}^{1}-S_{1})$ has a large component $C_{0}$(resp. $C_{1})$  and all the small components have at most $h-2$ vertices in total. Similar as the proof of Lemma \ref{fs2}, $C_{0}$ and $C_{1}$ are connected to each other in $Q_{n}-S$. So there exists a large component in $Q_{n}-S$.  Let $A$ be the union of  the vertex sets of all the small components in $Q_{n}-S$, then $|A|\leq 2(h-2)$. If $|A|\leq h-1$,  then we are done. Suppose not,  we have $h \leq |A|\leq 2(h-2)\leq 2(2n-5)=4n-10$.  Since $N_{Q_{n}}(A)\subset S$,  $|S|\geq |N_{Q_{n}}(A)|\geq b_{v}(|A|;Q_{n})$. According to the Lemma \ref{boundary2}, we have $b_{v}(h;Q_{n})\leq b_{v}(l;Q_{n})$ for $n+2\leq h \leq 2n-3$ and $h <l\leq 4n-10$.  So $|S|\geq b_{v}(|A|;Q_{n})\geq b_{v}(h;Q_{n})$,  this is a contradiction to $|S|<b_{v}(h;Q_{n})$.

 $\mathbf{Case~2.2}$. $|S_{0}|\geq b_{v}(h-1;Q_{n-1})$ or $|S_{1}|\geq b_{v}(h-1;Q_{n-1})$.

Without loss of generality,  we suppose $|S_{0}|\geq b_{v}(h-1;Q_{n-1})$. So $|S_{1}|\leq (h-2)\leq 2n-5< b_{v}(2;Q_{n-1})=2n-4$. Similarly as the proof of Case~1.2,   a contradiction can be obtained,   so  the result holds.\hfill\rule{1mm}{2mm}


\begin{lem}\label{fs4}
Let $S$ be a vertex set in $Q_{n}$  with $n\geq 7$ and $b_{v}(2n-3;Q_{n})\le |S|<b_{v}(2n+1;Q_{n})$,  then $Q_{n}-S$ has a large component and all the remaining small components have at most $2n$ vertices in total.
\end{lem}
$\mathbf{Proof:}$  Let $S_{i}=S\bigcap V(Q^{i}_{n-1})$,  $i=0,  1$. According  to Corollary 2.4,   $b_{v}(2n-2;Q_{n})= 3-3n+n^2$,  $b_{v}(2n+1;Q_{n})=-2 -2n +n^2$,  so $3-3n+n^2 \leq |S|\leq -3-2n+n^2$. Since $ 2 b_{v}(2n-5;Q_{n-1}) > |S|$ when $n\geq  6$,   so at most one of $|S_{0}|$ and $|S_{1}|$ can be not less than $ b_{v}(2n-5, Q_{n-1})$.

$\mathbf{Case ~1}$. $|S_{0}|< b_{v}(2n-5;Q_{n-1})$ and $|S_{1}|< b_{v}(2n-5;Q_{n-1})$.

By Lemma \ref{fs3},  $Q^{0}_{n-1}- S_{0}$ $(resp.Q^{1}_{n-1}- S_{1})$ has a large component $C_{0} (resp.C_{1})$,  and all the small components have at most $2n-6$ vertices in total. And similarly as the proof of Lemma $3.5$,   $C_{0}$ and $ C_{1}$ can be proved to be connected to each other in $Q_{n}-S$.  So there exists a large component $C$ in $Q_{n}-S$.   Let $A$ denotes the union of all the vertex sets of the small components in $Q_{n}-S$,  then $|A|\leq 4n-12 $. If $|A|\leq 2n $, then we are done. Suppose not, then $2n+1\leq  |A|< 4n-12$. Since $N_{Q_n}(A)\subset S$, $|S|\geq |N_{Q_n}(A)|$. But by Lemma \ref{boundary2}, $b_{v}(m, Q_n)\geq b_{v}(2n+1,Q_n)$ when $2n+1\leq m\leq 4n-12$. Thus $|S|\geq |N_{Q_n}(A)|\geq b_{v}(|A|,Q_n)\geq b_{v}(2n+1,Q_n) > |S|$, a contradiction. Thus the small components have at most $2n$ vertices in total.

 $\mathbf{Case ~2}$.  $|S_{0}|\geq b_{v}(2n-5 ;Q_{n-1})$ or $|S_{1}|\geq b_{v}(2n-5;Q_{n-1})$.

Without loss of generality,  we assume $|S_{1}|\geq  b_{v}(2n-5;Q_{n-1})$,  then $|S_0|\leq  -3-2n+n^2 - b_{v}(2n-5;Q_{n-1})= 3n-9<  b_{v}(3;Q_{n-1})$. So there eixists a large component in $Q_{n-1}^{0}-S_{0}$ and the small components have at most $2$ vertices in total . Since there exits a perfect matching between $Q_{n-1}^{0}$ and $Q_{n-1}^{1}$ in $Q_n$. At most $|S_0|+2$ vertices in $Q_{n-1}^{1} -S_1$ are not connected to the large component in  $Q_{n-1}^{0}-S_{0}$. So there exists a large component in $Q_n -S$ and the remaining small components have at most $(|S_0|+2)+2\leq 3n-5 $ vertices in total. Similar as the proof of case 1, we can prove that the small components have at most $2n$ vertices in total.
\hfill\rule{1mm}{2mm}

\begin{lem}\label{fs5}
Let $S$ be a vertex set in $Q_{n}$,  $ b_{v}(2n+1;Q_{n})\leq |S|< b_{v}(h;Q_{n})$ for $n\geq 9$ and $ 2n+2\leq h\leq 3n-6$,  then $Q_{n}-S$ has a large component and all the small components have at most $h-1$ vertices in total.
\end{lem}
$\mathbf{Proof:}$
 We use induction to prove this.

$\mathbf{(I)}$.When $h=2n+2$, Let $S_{i}=S\bigcap V(Q^{i}_{n-1})$,  $i=0,  1$. According  to Corollary 2.4,    $b_{v}(2n+1;Q_{n})= -2 -2n +n^2$ and $b_{v}(2n+2;Q_{n})= -7 -n +n^2$.  So $-2 -2n +n^2 \leq |S|\leq -8-n+n^2$. Since $ 2 b_{v}(2(n-1)+1;Q_{n-1}) > b_{v}(2n+2;Q_{n})$ when $n\geq  6$,   so at most one of $|S_{0}|$ and $|S_{1}|$ can be not less than $ b_{v}(2n-1, Q_{n-1})$.

$\mathbf{Case ~1.1}$. $|S_{0}|< b_{v}(2n-1;Q_{n-1})$ and $|S_{1}|< b_{v}(2n-1;Q_{n-1})$.

By Lemma \ref{fs4},  $Q^{0}_{n-1}- S_{0}$ $(resp.Q^{1}_{n-1}- S_{1})$ has a large component $C_{0} (resp.C_{1})$,  and all the small components have at most $2n-2$ vertices in total. And similarly as the proof of Lemma $3.5$,   $C_{0}$ and $ C_{1}$ can be proved to be connected to each other in $Q_{n}-S$.  So there exists a large component $C$ in $Q_{n}-S$.   Let $A$ denotes the union of all the vertex sets of the small components in $Q_{n}-S$,  then $|A|\leq 4n-4 $. If $|A|\leq 2n+1 $, then we are done. Suppose not, then $2n+2\leq  |A|< 4n-4$. Since $N_{Q_n}(A)\subset S$, $|S|\geq |N_{Q_n}(A)|$. But by Lemma \ref{boundary2}, $b_{v}(m, Q_n)\geq b_{v}(2n+2,Q_n)$ when $2n+2\leq m\leq 4n-4$. Thus $|S|\geq |N_{Q_n}(A)|\geq b_{v}(|A|,Q_n)\geq b_{v}(2n+1,Q_n) > |S|$, a contradiction. Thus the small components have at most $2n+1$ vertices in total.

 $\mathbf{Case ~1.2}$.  $|S_{0}|\geq b_{v}(2n-1 ;Q_{n-1})$ or $|S_{1}|\geq b_{v}(2n-1;Q_{n-1})$.

Without loss of generality,  we assume $|S_{1}|\geq  b_{v}(2n-1;Q_{n-1})$,  then $|S_0|\leq  -8-n+n^2- b_{v}(2n-1;Q_{n-1})= 3n-9<  b_{v}(3;Q_{n-1})$. So there eixists a large component in $Q_{n-1}^{0}-S_{0}$ and the small components have at most $2$ vertices in total . Since there exits a perfect matching between $Q_{n-1}^{0}$ and $Q_{n-1}^{1}$ in $Q_n$. At most $|S_0|+2$ vertices in $Q_{n-1}^{1} -S_1$ are not connected to the large component in  $Q_{n-1}^{0}-S_{0}$. So there exists a large component in $Q_n -S$ and the remaining small components have at most $(|S_0|+2)+2\leq 3n-5 $ vertices in total. Similar as the proof of case 1, we can prove the small components have at most $2n+1$ vertices in total.

$\mathbf{(II)}$. Suppose the proposition holds for $h-1$ where $3n-5\geq h\geq 2n+3$.  In the following we will prove it also holds for $h$. Suppose not,  let $C_{1}, C_{2}, \cdots, C_{m}$ be all the components in $Q_{n}-S$ with  $|V(C_{1})|\leq |V( C_{2})|\leq \cdots \leq V(| C_{m}|)$;  that is,  $m\geq 2$  and $\sum_{i=1}^{m-1}|V(C_{i})|\geq h$.  Let $ B=\bigcup _{i=1}^{m-1}V(C_{i}) $, $u, v$ be two different vertices in $B$. Suppose they differ in the $i$-th dimension,
  let $Q_{n}=G(Q^{0}_{n-1}, Q^{1}_{n-1};M)$ be a decomposition of $Q_{n}$ along the $i$-th dimension.  Then $u$ and $ v$ don't belong to the same subcube.

Since $b_{v}(h-1;Q_{n-1})+ (2h-3n+1)=b_{v}(h;Q_{n})$,  it can be verified that $2h-3n+1\leq b_{v}(h-1;Q_{n-1}) $ when $ n\geq 7$ and $h\geq 2n+3$. So at most one of $|S_{0}|$ and $|S_{1}|$ can be not  less than $ b_{v}(h-1;Q_{n-1})$.

$\mathbf{Case ~2.1}$. $|S_{0}|<b_{v}(h-1;Q_{n-1})$ and $|S_{1}|< b_{v}(h-1;Q_{n-1})$.

By the induction hypothesis, $Q_{n-1}^{0}-S_{0}$(resp. $Q_{n-1}^{1}-S_{1})$ has a large component $C_{0}$(resp. $C_{1})$  and all the small components have at most $h-2$ vertices in total. Similar as the proof of Lemma \ref{fs2}, $C_{0}$ and $C_{1}$ are connected to each other in $Q_{n}-S$. So there exists a large component in $Q_{n}-S$.  Let $A$ be the union of  the vertex sets of all the small components in $Q_{n}-S$, then $|A|\leq 2(h-2)$. If $|A|\leq h-1$,  then we are done. Suppose not,  we have $h \leq |A|\leq 2(h-2)\leq 2(3n-8)=6n-16$.  Since $N_{Q_{n}}(A)\subset S$,  $|S|\geq |N_{Q_{n}}(A)|\geq b_{v}(|A|;Q_{n})$. According to the Lemma \ref{boundary2}, we have $b_{v}(h;Q_{n})< b_{v}(l;Q_{n})$ for $2n+2\leq h \leq 3n-6$ and $h <l\leq 6n-16$.  So $|S|\geq b_{v}(|A|;Q_{n})> b_{v}(h;Q_{n})$,  this is a contradiction to $|S|<b_{v}(h;Q_{n})$.

 $\mathbf{Case~2.2}$. $|S_{0}|\geq b_{v}(h-1;Q_{n-1})$ or $|S_{1}|\geq b_{v}(h-1;Q_{n-1})$.

Without loss of generality,  we suppose $|S_{0}|\geq b_{v}(h-1;Q_{n-1})$. So $|S_{1}|\leq 2h-3n+1\leq 3n-11< b_{v}(3;Q_{n-1})=3n-8$. Similarly as the proof of Case~1.2,   a contradiction can be obtained,   so  the result holds. \hfill\rule{1mm}{2mm}

Let $n$, $h\in N^+$,  $1\le h \le 3n-6$, we define $f(h)$ as follows:

\begin{displaymath}
f(h) =\left\{\begin{array}{l l}
h-1 ,    & \textrm{ $1\leq h \leq n-2$}\\
n+1 ,     & \textrm{$ h=n-1,n $} \\
n,   & \textrm{$h=n+1$ } \\
h-1,    & \textrm{$n+2 \leq h  \leq  2n-3$ }\\
2n,    & \textrm{$ h=2n-2,2n-1,2n+1 $}\\
2n-4,    & \textrm{$ h=2n $}\\
h-1.   & \textrm{$ 2n+2 \leq  h \leq 3n-6 $}
\end{array} \right.
\end{displaymath}

By Lemma \ref{fs},  Lemma \ref{fs2}, Lemma \ref{fs3},  Lemma \ref{fs4} and Lemma \ref{fs5},  the following Theorem can be obtained:
\begin{thm}\label{$Q_{n-1}^{0}`}

Given an $n$-dimensioanl hypercube $Q_n$ , for any vertex subset  $S$ of $Q_{n}$ with  $|S|< b_{v}(h;Q_{n})$ with  $1 \le h\leq 3n-6$,  then
 there exists a large component in $Q_{n}-S$ and all the remaining small components have at most $f(h)$ vertices in total.

\end{thm}

The following theorem shows the relationship between the hypercube's
extra connectivity and its minimum boundary number.
\begin{thm}
Let $Q_n$ be an $n$-dimensional hypercube and $1\le h \le 3n-6$,   the $h-1$-extra connectivity of $Q_{n}$ are  as follows :

\begin{displaymath}
\kappa_{h-1}(Q_{n}) =\left\{\begin{array}{ll}
 b_{v}(h, Q_{n}) ,    & \textrm{ $1\le h \leq n-3, n\ge 5$}\\
b_{v}(n-2, Q_{n}) ,    & \textrm{$ n-2\leq h \leq n+1, n\ge 5$ }\\
b_{v}(h, Q_{n}) .   & \textrm{$ n+2 \leq h \leq 2n-4, n\ge 7 $ }\\
b_{v}(2n-3, Q_{n}) ,    & \textrm{$ 2n-3\leq h \leq 2n, n\ge 7$ }\\
b_{v}(h, Q_{n}) .   & \textrm{$ 2n+1 \leq h \leq 3n-6, n\ge9 $ }
\end{array} \right.
\end{displaymath}

\end{thm}

$\mathbf{Proof:}$
1) Let $Q_n$ be a $n$ dimensional  Hypercube with   $n\ge 5$. When   $h\le n-2$,  by Lemma \ref{fs},   if $|S|<b_v(h;Q_n)$,  there exists   a large component    in $Q_n-S$  and  all  the small components have at most $h-1$ vertices in total.  Thus  $\kappa_{h-1}(Q_n)\ge b_v(h;Q_n)$ .

Let $u=0^n$, $S_h= \{u, u^{1}, \cdots u^{h-1}\}$, when $1\le h \le n+1$. By Lemma \ref{cnnumber} and Lemma \ref{boundary}, $|N_{Q_n}(S_h)|= (n-h+1)+ (n-1)(h-1)-\binom{h-1}{2}= b_{v}(h, Q_{n})$. It's clear that $[S_h]$ is isomorphic to $K_{1,h-1}$.  When $1\le h \le n-3$,  by Lemma \ref{boundary2} $b_{v}(h, Q_{n})<b_{v}(h+1, Q_{n})$.  Thus according to  Lemma \ref{fs}, there exists a large component in $Q_n - N_{Q_n}(S_h)$ and the small components have at most $h$ vertices in total. Obviously $[S_h]$ is a connected component in $Q_n - N_{Q_n}(S_h)$ with $h$ vertices. So $Q_n - C_{Q_n}(S_h)$ is the large connected component in $Q_n - N_{Q_n}(S_h)$.  $N_{Q_n}(S_h)$ is an $(h-1)$-extra vertex cut of $Q_n$. Thus $\kappa_{h-1}(Q_n)\le |N_{Q_n}(S_h)|= b_v(h; Q_n)$ when $1\le h \le n-3$.

2) Let $u=0^n$, $S_n= \{u, u^{1}, \cdots u^{n}\}$.
By Lemma \ref{boundary2}, $b_v(n-2; Q_n)+1=b_v(n+1; Q_n)+1=b_v(n-1; Q_n)=b_v(n; Q_n)$, so  $|N_{Q_n}(S_{n})|= b_v(n+1; Q_n)<b_v(n-1;Q_n)$.  According to Lemma \ref{fs2},  when $n\ge 5$,  $Q_n - N_{Q_n}(S_{n})$ has a large connected component and all the small components have at most $n+1$ vertices in total. It's clear that $[S_{n}]$ is a connected component in $Q_n - N_{Q_n}(S_{n})$ with $n+1$ vertices.  So  $Q_n - C_{Q_n}(S_{n})$ is connected. Thus  $N_{Q_n}(S_{n})$ is an $n$-extra vertex cut,  $\kappa_{n}(Q_n)\le |N_{Q_n}(S_{n})|= b_v(n-2; Q_n) $.

Let $S$ be an $(n-3)$-extra vertex cut of $Q_n$, then each component of $Q_n -S$ has at least $n-2$ vertices. By Lemma \ref{fs}, $|S|\ge b_v(n-2; Q_n)$. Thus $\kappa_{n-3}(Q_n)\ge b_v(n-2; Q_n)$.

So we have $\kappa_{n-3}(Q_n)\ge b_v(n-2; Q_n)\ge \kappa_{n}(Q_n)\ge \kappa_{n-3}(Q_n)$. The inequalities are all equalities. So $\kappa_{h-1}(Q_n)= b_v(n-2; Q_n)$ for $n-2 \le h \le n+1$.

3) By Lemma  \ref{fs3},  $\kappa_{h-1}(Q_n)\ge b_v(h;Q_n)$ when $n\ge 7$, $n+2\le h\le 2n-3$. In the following paragraph, we will prove that  $\kappa_{h-1}(Q_n)\le b_v(h; Q_n)$ when $n+2 \le h \le 2n-4$.

Let $u=0^n$, Let $S_h= \{u, u^{1}, \cdots u^{n}, u^{12}, u^{13},  \cdots u^{1(h-n)}\}$,when $n+2 \le h \le 2n $ .  Then it's easy to verify that $|N_{Q_n}(S_h)|= b_{v}(h, Q_{n})$. It's clear that $[S_h]$ is connected.  When $n+2\le h \le 2n-4$, by Lemma \ref{boundary2}, $b_{v}(h, Q_{n})<b_{v}(h+1, Q_{n})$.  Thus by Lemma \ref{fs2} there exists a large component in $Q_n - N_{Q_n}(S_h)$ and the small components have at most $h$ vertices in total. Obviously $[S_h]$ is a connected component in $Q_n - N_{Q_n}(S_h)$ with $h$ vertices, so $Q_n - C_{Q_n}(S_h)$ is the large connected component in $Q_n - N_{Q_n}(S_h)$. So $N_{Q_n}(S_h)$ is an $h-1$-extra vertex cut of $Q_n$. Thus $\kappa_{h-1}(Q_n)\le b_v(h; Q_n)$ when $n+2 \le h \le 2n-4$.

4) Let $u=0^n$,  $S_{2n-1}= \{u, u^{1}, \cdots u^{n}, u^{12}, u^{13},  \cdots u^{1n}\}$.
By Lemma \ref{boundary2}, $b_v(2n-3; Q_n)+1=b_v(2n; Q_n)+1=b_v(2n-2; Q_n)=b_v(2n-1; Q_n)$. Since $|N_{Q_n}(S_{2n-1})|= b_v(2n; Q_n)<b_v(2n+1;Q_n)$,  according to Lemma \ref{fs4}, $Q_n - N_{Q_n}(S_{2n-1})$ has a large connected component and all the small components have at most $2n$ vertices in total. It's clear that $[S_{2n-1}]$ is a connected component in $Q_n - N_{Q_n}(S_{2n-1})$ with $2n$ vertices.  So  $Q_n - C_{Q_n}(S_{2n-1})$ is connected. Thus  $N_{Q_n}(S_{2n-1})$ is an $(2n-1)$-extra vertex cut. Thus $\kappa_{2n-1}(Q_n)\le |N_{Q_n}(S_{2n-1})|= b_v(2n; Q_n) $.

Let $S$ be a  minimum $(2n-4)$-extra vertex cut of $Q_n$, then each component of $Q_n -S$ has at least $2n-3$ vertices. By Lemma \ref{fs4}, $|S|\ge b_v(2n-3; Q_n)$. Thus $\kappa_{2n-4}(Q_n)\ge b_v(2n-3; Q_n)$.

So we have $b_v(2n-3; Q_n)\le \kappa_{2n-4}(Q_n)\le \kappa_{2n-3}(Q_n)\le \kappa_{2n-2}(Q_n)\le \kappa_{2n-1}(Q_n) \le b_v(2n; Q_n)$.  Since $b_v(2n-3; Q_n)=b_v(2n; Q_n)$, the inequalities are all equalities. So $\kappa_{h-1}(Q_n)= b_v(2n-3; Q_n)$ for $2n-3 \le h \le 2n$.

5) By  Lemma \ref{fs5},  $\kappa_{h-1}(Q_n)\ge b_v(h; Q_{n})$ when $n\ge 9$, $2n+1\le h\le 3n-6$.

For an integer $2n+1\le h\le 3n-6$, let  $u=0^n$ and $k= h+1-2n$.  Let $S_h= \{u, u^{1}, \cdots u^{n}, u^{12}, \cdots u^{1n}, u^{2n}, u^{3n}, \cdots u^{kn}\}$ with $2 \le k \le n-1 $, when $2n+1 \le h \le 3n-2$.  By Lemma \ref{cnnumber} and Corollary \ref{boundary}, it's easy to verify that $|N_{Q_n}(S_h)|= b_{v}(h, Q_{n})$. It's clear that $[S_h]$ is connected.  When $2n+1\le h \le 3n-6$, by Lemma \ref{boundary2},  $b_{v}(h, Q_{n})<b_{v}(h+1, Q_{n})$.  Thus by Lemma \ref{fs5} there exists a large component in $Q_n - N_{Q_n}(S_h)$ and the small components have at most $h$ vertices in total. Obviously $[S_h]$ is a connected component in $Q_n - N_{Q_n}(S_h)$ with $h$ vertices, so $Q_n - C_{Q_n}(S_h)$ is the large connected component in $Q_n - N_{Q_n}(S_h)$. So $N_{Q_n}(S_h)$ is an $(h-1)$-extra vertex cut of $Q_n$. Thus $\kappa_{h-1}(Q_n)\le b_v(h; Q_n)$ when $2n+1 \le h \le 3n-6$.
\hfill\rule{1mm}{2mm}

\section{Conclusion}
We have proposed a new approach to finding a network's conditional connectivity based on its isoperimetric properties.
Using the vertex isoperimetric results, we have analyzed a faulty hypercube $Q_n$'s structure, and determined its
$h$-extra connectivity for $1\le h\le 3n-6$.

Our work in this paper is the first attempt to establish a link between the two fields,
i.e. between graph isoperimetric problems and connectivity/reliability of interconnection networks.
We have shown that the results and methods in the former can be applied in the study of the latter.
We anticipate that the established link will help getting more insights and expanding toolkits
for the research of interconnection networks.

%

%
%
%

%

\end{document}